\documentclass{ws-procs9x6}
\newcommand{\bce}{\begin{center}}
\newcommand{\ece}{\end{center}}
\newcommand{\be}{\begin{equation}}
\newcommand{\ee}{\end{equation}}
\newcommand{\bea}{\begin{eqnarray}}
\newcommand{\eea}{\end{eqnarray}}
\newcommand{\non}{\nonumber \\}
\newcommand{\no}{\nonumber}
\def\E{\> = \>}
\def\EA{&=&}
\newcommand{\bdes}{\begin{description}}
\newcommand{\edes}{\end{description}}
\newcommand{\bit}{\begin{itemize}}
\newcommand{\eit}{\end{itemize}}
\newcommand{\bsm}{\begin{small}}
\newcommand{\esm}{\end{small}}
\newcommand{\bfoot}{\begin{footnotesize}}
\newcommand{\efoot}{\end{footnotesize}}
\newcommand{\bscr}{\begin{scriptsize}}
\newcommand{\escr}{\end{scriptsize}}
\newcommand{\btin}{\begin{tiny}}
\newcommand{\etin}{\end{tiny}}
\newcommand{\deF}{\> = \, : \>}
\newcommand{\Def}{\> : \, = \>}

\newcommand{\fk}{{\bf k}}
\newcommand{\fx}{{\bf x}}

\begin{document}

\title{PERTURBATIVE RESULTS WITHOUT DIAGRAMS}

\author{R. ROSENFELDER}

\address{Particle Theory Group, Laboratory for Particle Physics \\
Paul Scherrer Institut\\
CH-5232 Villigen PSI, Switzerland\\
E-mail: roland.rosenfelder@psi.ch}

\begin{abstract}
Higher-order perturbative calculations in Quantum
(Field) Theory suffer from the factorial increase of the number of
individual diagrams. Here I describe an approach which evaluates
the total contribution numerically for finite temperature
from the cumulant expansion of the corresponding observable followed
by an extrapolation to zero temperature.
This method (originally proposed by Bogolyubov and Plechko) is applied
to the calculation of higher-order terms for the
ground-state energy of the polaron. Using
state-of-the-art multidimensional integration routines 2 new coefficients 
are obtained corresponding to a 4- and 5-loop
calculation.
\end{abstract}

\keywords{high-order perturbative calculations, cumulant expansion, Monte-Carlo integration}

\bodymatter

\section{Introduction}
Highly accurate measurements require precise theoretical calculations which perturbation theory
can yield if the coupling constant is small. However, in Quantum Field Theory (QFT)
the number of diagrams grows factorially with the order of perturbation theory and they become more
and more complicated. The prime example is the anomalous magnetic moment of the electron where
new experiments$^{\mbox{\refcite{Gab}}}$ need high-order quantum-electrodynamical calculations but 
the number of diagrams for them ``explodes'' as shown by the generating 
function$^{\mbox{\refcite{ItZu}}}$
\be
\Gamma(\alpha) \E 1 + \alpha + 7 \, \alpha^2 + 72 \, \alpha^3 + 891 \, \alpha^4 + 12672 \, \alpha^5 + 
202770 \, \alpha^6 
+ \ldots
\ee
There are ongoing efforts$^{\mbox{\refcite{Kino}}}$ to calculate all $12672$ diagrams 
in ${\cal O}(\alpha^5)$ -- a huge, heroic effort considering 
the complexity of individual diagrams and the large cancellations among them.

Obviously new and more efficient methods would be most welcome for a cross-check 
or further progress.

\section{A new method (applied to the polaron g.s. energy)}
Here I present a ``new'' method which  -- as I learned during the conference -- 
was already proposed 20 years by Bogolyubov (Jr.) and Plechko (BP)~$^{\mbox{\refcite{BoPl}}}$. 
However, to my knowledge it has been never applied numerically which turned out to be quite a 
challenging  task.

The BP method is formulated for the polaron problem, a non-relativistic (but non-trivial) 
field theory describing an electron slowly moving through a polarizable crystal. Due to
medium effects its energy is changed and it acquires an effective mass : 
$ E_{\bf p} = E_0 + {\bf p}^2/(2 m^{\star}) + \ldots $.
The aim is to calculate the power series expansion for the g.s. energy 
\be
E_0(\alpha) \deF \sum_{n=1} e_n \, \alpha^n 
\ee
as function of the dimensionless electron-phonon coupling constant $\alpha$.
The lowest-order coefficients are well-known$^{\mbox{\refcite{HoMu}}}$
( $ e_1 \E - 1 \>  , \hspace{0.2cm} e_2 \E -0.015 91962 $~) but since Smondyrev's 
calculation$^{\mbox{\refcite{Smon}}}$ in 1986
\be
e_3 \E -0.000 80607  
\label{e3}
\ee
there has been no progress towards higher-order terms.

This will be remedied by the first numerical application of the BP method. For this purpose
the path integral formulation of the polaron problem will be used where the phonons have 
been integrated out exactly$^{\mbox{\refcite{Feyn}}}$. For large Euclidean times $\beta$ 
this gives the following effective action
\be
\hspace*{-0.2cm} S_{\rm eff}[\fx] \!=\! \int_0^{\beta} \! \!dt  \frac{1}{2} \dot
 \fx^2 -  \frac{\alpha}{\sqrt{2}} 
\int_0^{\beta} \! \! dt \!\int_0^t \! \! dt' \, e^{-(t-t')} \!\int \!\frac{d^3k}{2 \pi^2}  \, 
\frac{\exp \left [ i {\fk} \cdot \left ( {\fx}(t) - {\fx}(t') \right) \right ]}{{\fk}^2} 
\> 
\ee
which will be split into a free part $ S_0$ and an interaction term $S_1$.
The g.s. energy may be obtained from the partition function
\be
Z(\beta) \E \oint {\cal D}^3 x \> e^{-S_{\rm eff}[\fx]} \> \>
\stackrel{\beta \to \infty}{\longrightarrow} \> \>   e^{-\beta E_0} 
\ee
at asymptotic values of $\beta$, i.e. zero temperature. 
The central idea is to use the {\it cumulant expansion} of the partition function
\be
Z(\beta) \E Z_0 \, \exp \left [  \sum_{n=1} \frac{(-)^n}{n !} \lambda_n(\beta)
 \right] 
\ee
where the $ \lambda_n(\beta) $'s are the cumulants w.r.t. $ S_1 $. These are obtained
from the {\it moments}
\be
m_n \Def  {\cal N}  \, \oint {\cal D}^3x \>
 \left ( \, S_1[x] \, \right )^n \> e^{-S_0[x]} \> \> \> , \> \> \> m_0 \E 1 
\ee
by the recursion relation (see, e.g. Eq. (51) in Ref.~\refcite{quasi})
\be
\lambda_{n+1} \E m_{n+1}  -  \sum_{k=0}^{n-1}
\binom{n}{k} \lambda_{k+1}  m_{n-k}  \> .
\label{cum recurs}
\ee
Explicitly the first cumulants read
\bea
\lambda_1 \EA m_1  \> \> , \> \> \lambda_2 \E m_2 -  m_1^2 \> \> , \> \> 
\lambda_3 \E m_3 - 3 \, m_2 \, m_1 + 2 \, m_1^3  \non
\lambda_4 \EA m_4 - 4 \, m_3 \, m_1 - 3 \, m_2^2 + 12 \, m_2\,  m_1^2 -
 6 \, m_1^4  
\label{cum} \\
\lambda_5 \EA m_5 - 5 \, m_4 \, m_1 - 10 \, m_3 \, m_2 + 20 \, m_3 \, m_1^2 +
30 \, m_2^2 \, m_1 - 60 \, m_2 \, m_1^3 + 24 \, m_1^5 \no
\eea
By construction $ \> m_n \propto \alpha^n \> $ and Eq. (\ref{cum recurs})
shows that the cumulants share this property. Thus we immediately obtain 
\be
e_n \E \lim_{\beta \to \infty} \frac{1}{\beta} \,
\frac{(-)^{n+1}}{\alpha^n \, n !} \, \lambda_n(\beta)  \> . 
\label{en limit}
\ee
The functional integral for the moments can be done since it is Gaussian. 
The integrals over the phonon momenta $\fk_m \> , \> m = 1 \ldots n $ can also be performed
if the $m^{\rm th}$ propagator is written as
\be
\frac{1}{\fk_m^2} \E \frac{1}{2} \,
\int_0^{\infty} du_m \> \exp \left [ - \frac{1}{2} \fk_m^2 \, u_m \right ] \>.
\ee
Then one obtains
\bea
m_n \EA \frac{(-)^n\alpha^n}{(4 \pi)^{n/2}} \, \prod_{m=1}^n \left ( \int_0^{\beta} \!
dt_m \int_0^{t_m}\! dt_m' \int_0^{\infty} \! du_m \right ) \, \exp \left [ -
 \sum_{m=1}^n(t_m - t_m') \right ] \non
&& \hspace{3cm} \cdot  \left [ \, \det A \left (t_1 \ldots  t_n, t_1'
\ldots  t_n';
u_1 \ldots u_n \right ) \, \right ]^{-3/2} \> .
\label{m_n}
\eea
Here the $(n \times n)$- matrix $A$ 
\be
A_{i j} \E \frac{1}{2}  \> \Bigl [ \> \, - |t_i -t_j| + |t_i - t_j'| +
| t_i'- t_j| -|t_i' - t_j'| \>\,   \Bigr ] + u_i \, \delta_{ij} \> .
\label{det}
\ee
is non-analytic in the times $t_i, t_i'$,  but analytic
in the auxiliary variables $u_i$.

\section{Numerical procedures and results}
The task is now to perform the $(3n)$-dimensional integral over $t_i,t_i',u_i$ for large 
enough $\beta$ in the expression for the cumulants/moments.
It is clear that any reduction in the dimensionality of the integral
will greatly help in obtaining reliable numerical results in affordable CPU-time. A closer 
inspection of the structure of the integrand reveals that 2 integrations over the auxiliary 
variables (say $u_n, u_{n-1}$) can always be done analytically. Furthermore, 
we do not use Eq. (\ref{en limit}) to extract the energy coefficient $e_n$ but 
\be
e_n  \E   \frac{(-)^{n+1}}{ \alpha^n n !}\lim_{\beta \to \infty} \, 
\frac{\partial \lambda_n (\beta)}{\partial \beta} \deF \lim_{\beta \to \infty} \,e_n(\beta) \> .
\label{en diff}
\ee
This ``kills two birds with one stone'': first the derivative w.r.t. $\beta$ takes away one further
integration over a time (see Eq. (\ref{m_n}) where $\beta$ appears as upper limit) requiring that 
only a $(3n-3)$-dimensional integral has to be done numerically. Second, it vastly improves the 
convergence to $e_n \equiv e_n(\beta = \infty)$ because now
\be
e_n(\beta)  \> \stackrel{\beta \to \infty}{\longrightarrow} \>  \frac{\partial}{\partial \beta} 
\, \Bigl [ \, \beta\, \cdot \ e_n +
{\rm const} - \frac{a_n}{\sqrt{\beta}} \, e^{-\beta} + \ldots \,  \Bigr ]
\E e_n +  \frac{a_n}{\sqrt{\beta}} \, e^{-\beta} + \ldots \> .
\label{en asy}
\ee
In other words : we obtain an {\it exponential} convergence to the value $e_n$ 
whereas previously the approach would be very slow, 
like ${\rm const}/\beta$. This exponential convergence of the derivative version
has been demonstrated analytically for $n = 1, 2$ and numerically for $n = 3$ (see below). In the 
following we will assume that it holds for {\it all} $n$.
After mapping to the hypercube $[0,1]$  the remaining $(3n-3)$-dimensional integral can be 
evaluated by Monte-Carlo techniques utilizing the classic
VEGAS program$^{\mbox{\refcite{Vegas}}}$
or the more modern programs from the CUBA library$^{\mbox{\refcite{Cuba}}}$. 

We first have tested this approach by comparing with the analytical result
\begin{figure}
\bce
\psfig{file=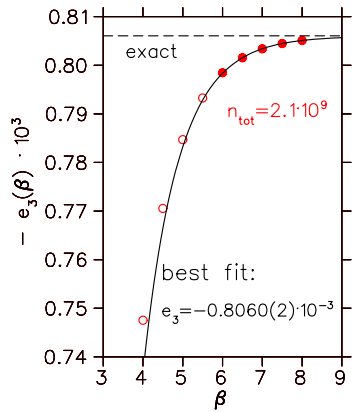,width=45mm}
\ece
\caption{(color online). Monte-Carlo results for the derivative of the $3^{\rm rd}$ cumulant as
function of the Euclidean time $\beta$. The total number of function calls is denoted 
by $n_{\rm tot}$ 
and the full (open) circles are the points used (not used) in the fit.}
\label{fig1}
\end{figure}
given in Eq. (\ref{e3}). Fig. \ref{fig1} shows $e_3(\beta)$ 
and the best fit to the data assuming the $\beta$-dependence (\ref{en asy}). 
Since the asymptotic behaviour is not valid for low values 
of $\beta$ we have eliminated small-$\beta$ points successively until the resulting $\chi^2$/dof 
of the fit reaches a minimum. Excellent agreement with Smondyrev's result (\ref{e3}) is found.
If one allows for a different power of $\beta$ in the prefactor
of Eq. (\ref{en asy}) then the fit gives an exponent $ \> -0.55 (3) \> $ instead of $-0.5$ assumed 
before.

\begin{figure}
\begin{tabular}[b]{@{}l@{}}
  \psfig{file=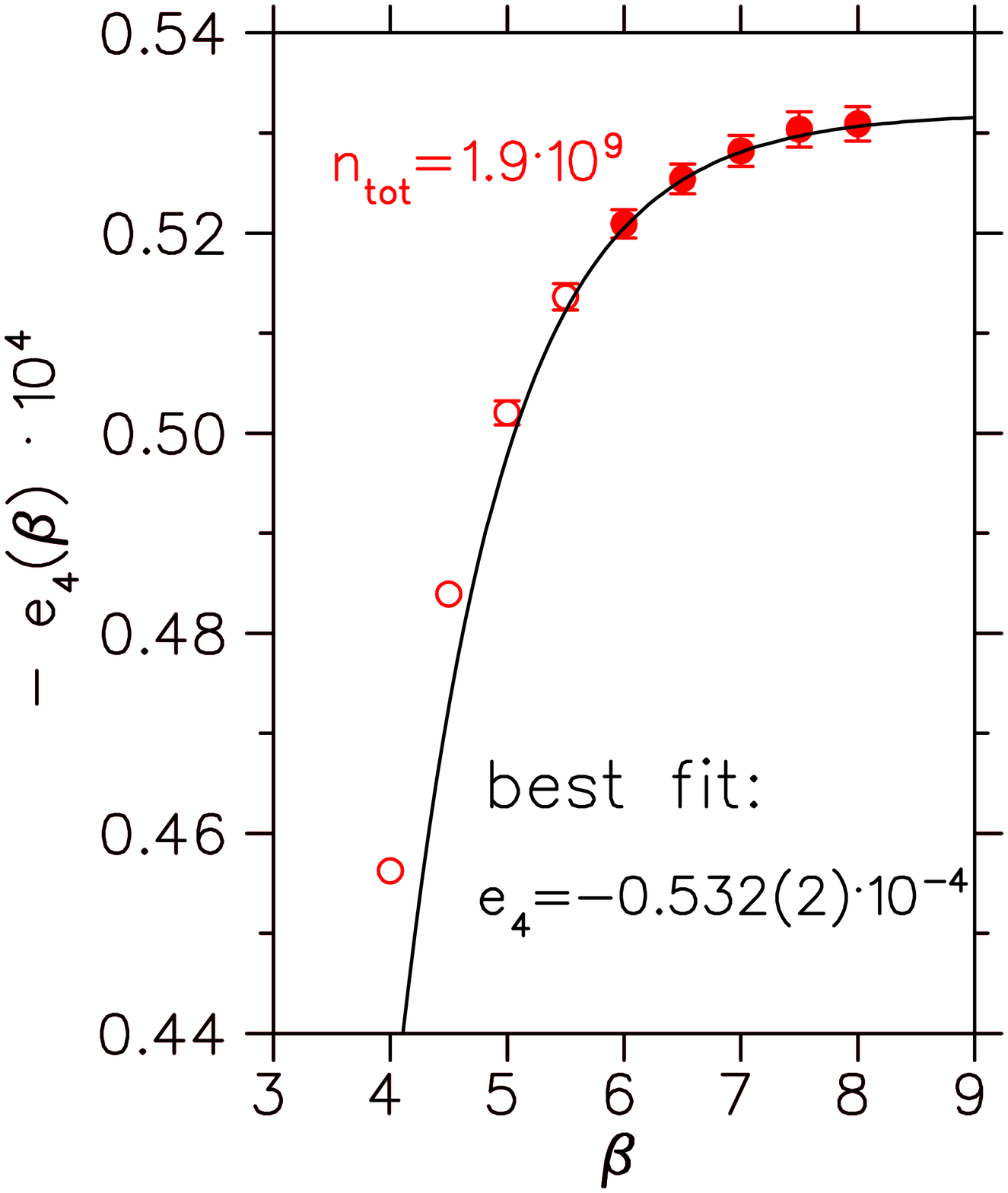,width=45mm} \\
  \hspace*{3cm}{\bf (a)}
\end{tabular}
\hfill
\begin{tabular}[b]{@{}c@{}}
  \psfig{file=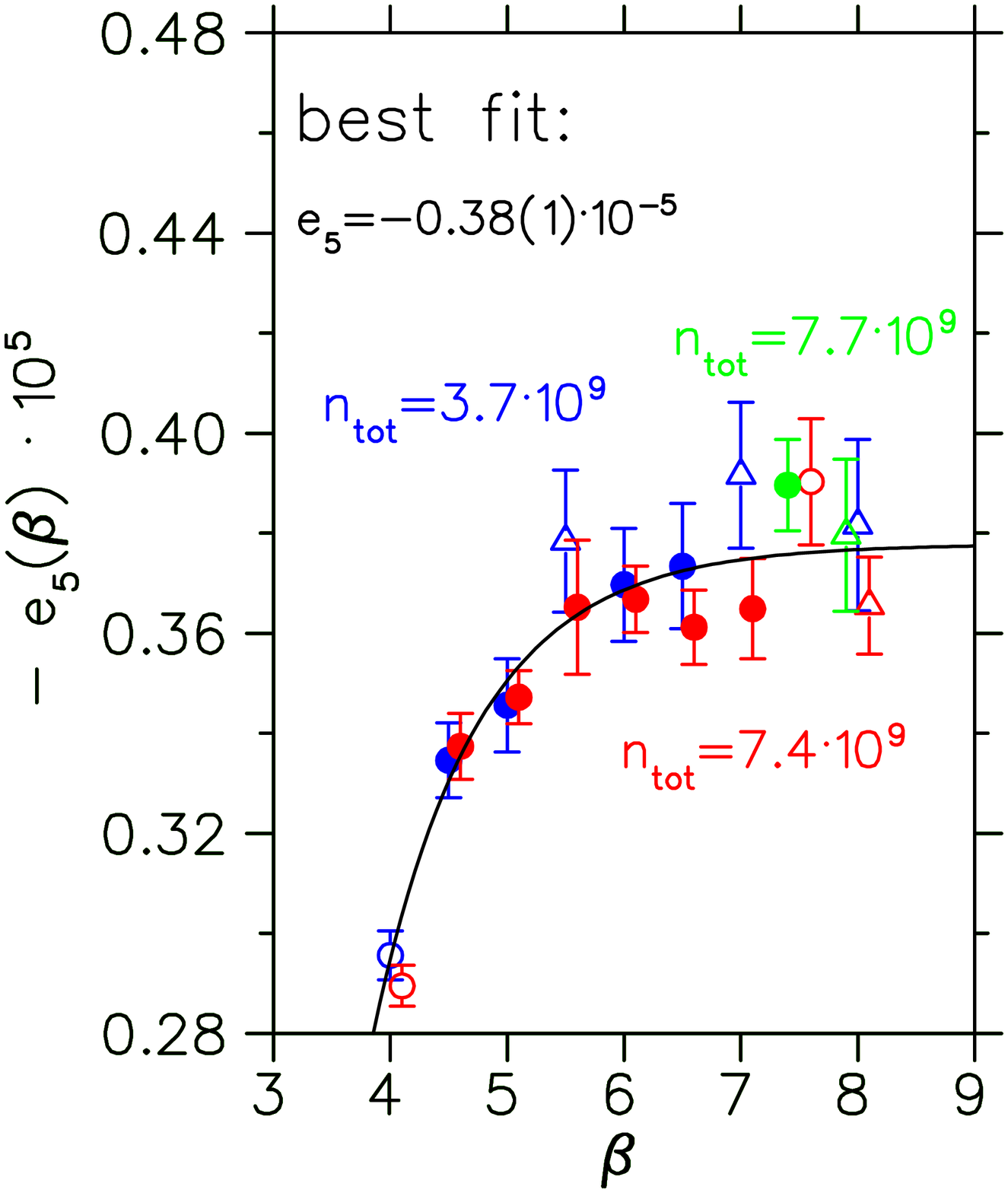,width=45mm} \\
 \hspace*{1cm} {\bf (b)}
\end{tabular}
\caption{(a) Same as Fig. \ref{fig1} but for the $4^{\rm th}$ cumulant. (b) Data for the 
derivative of the $5^{\rm th}$ cumulant. Open triangles denote results (not used in the fit)
which have a 
$\chi^2 > 1.5$ indicating that successive Monte-Carlo iterations 
are not consistent with each other.}
\label{fig2}
\end{figure}

However, when extending these calculations to the case $ n = 4 $ a very slow convergence
of the numerical result with the number of function calls $ n_{\rm tot} $ is observed 
at fixed $\beta$.
Fortunately, a solution was found by performing
the remaining $ (n-2) \> u_i$-integrations not by stochastic (Monte-Carlo) methods but by
deterministic quadrature rules. This is possible since the $u_i$-dependence of the integrand
is analytic (see Eq. (\ref{det})). We have used the very efficient ``tanhsinh-integration'' 
method$^{\mbox{\refcite{tanhsinh}}}$ but Gaussian quadrature is nearly as good. 
A dramatic improvement in stability results together 
with a reduction of $ n_{\rm tot} $ needed
for the much smaller values of $ |e_n| \> , n > 3 $.
This allows a reliable evaluation of $e_4$ (see Fig. \ref{fig2} a) and also makes the determination of
$e_5$ feasible as shown in Fig. \ref{fig2} b. 

\noindent
The best fit values for $e_4$ and $e_5$ displayed in Figs. \ref{fig2} a, b 
are still preliminary as a more detailed error analysis has to be made. Also for the
$n = 5$ case the Monte-Carlo statistics should be improved. Note that each   
high-statistic point in Fig. \ref{fig2}~b 
took about 30 days runtime on a Xeon 3.0 GHz machine. 

\section{Summary and Outlook}

\bit
\item Two additional perturbative coefficients $e_4, e_5 $ for the polaron g.s.
energy have been determined by the method of Bogolyubov and Plechkov (rediscovered
independently). This
amounts to performing a 4-loop and 5-loop calculation in Quantum Field Theory.

\item The method is based on a combination of Monte-Carlo integration techniques and 
deterministic quadrature rules for finite $\beta$ (temperature) and on a 
judicious extrapolation to $ \beta \to \infty $ (zero temperature). 
As a check the value of $e_3$ calculated analytically by Smondyrev has been reproduced 
with high accuracy.

\item The cancellation in $n^{\rm th}$ order is {\it  not} among many individual diagrams
but among the much fewer terms in the integrand of the $(3n-3)$-dimensional integral
(see Eq. (\ref{cum})).


\item The method can be simply extended to the calculation of higher-order terms in the
small-coupling expansion of the effective mass $m^{\star}(\alpha)$ for a moving polaron.

\item Generalizing this approach to relativistic QFT in the 
worldline representation$^{\mbox{\refcite{worldline}}}$ and calculation of higher-order terms 
for the anomalous 
magnetic moment of the electron is under investigation. New challenges arise 
from the divergences which now occur and the need for renormalization.
\eit

\end{document}